\documentclass[prl,twocolumn,superscriptaddress,longbibliography]{revtex4-2}
\usepackage{graphicx,amsmath,amssymb,tikz,scalerel,mathrsfs,hyperref,comment}
\usepackage{soul}
\usepackage{color, xcolor}
\hypersetup{
        colorlinks=true,       
        linkcolor=red,        
        citecolor=blue,     
        urlcolor=blue}
\renewcommand{\section}[1]{{\par\it #1.---}\ignorespaces}
\usetikzlibrary{svg.path}
\definecolor{orcidlogocol}{HTML}{A6CE39}
\tikzset{
	orcidlogo/.pic={
		\fill[orcidlogocol] svg{M256,128c0,70.7-57.3,128-128,128C57.3,256,0,198.7,0,128C0,57.3,57.3,0,128,0C198.7,0,256,57.3,256,128z};
		\fill[white] svg{M86.3,186.2H70.9V79.1h15.4v48.4V186.2z}
		svg{M108.9,79.1h41.6c39.6,0,57,28.3,57,53.6c0,27.5-21.5,53.6-56.8,53.6h-41.8V79.1z M124.3,172.4h24.5c34.9,0,42.9-26.5,42.9-39.7c0-21.5-13.7-39.7-43.7-39.7h-23.7V172.4z}
		svg{M88.7,56.8c0,5.5-4.5,10.1-10.1,10.1c-5.6,0-10.1-4.6-10.1-10.1c0-5.6,4.5-10.1,10.1-10.1C84.2,46.7,88.7,51.3,88.7,56.8z};}}
\newcommand\orcid[1]{\href{https://orcid.org/#1}{\mbox{\scalerel*{\begin{tikzpicture}[yscale=-1,transform shape]\pic{orcidlogo};\end{tikzpicture}}{|}}}}

\begin{document}
\title{Remote Charging and Degradation Suppression for the Quantum Battery}
\author{Wan-Lu Song\orcid{0000-0002-6437-9748}}
\affiliation{Department of Physics, Hubei University, Wuhan 430062, China}
\author{Hai-Bin Liu}
\affiliation{Department of Physics, Hubei University, Wuhan 430062, China}
\author{Bin Zhou\orcid{0000-0002-4808-0439}}
\email{binzhou@hubu.edu.cn}
\affiliation{Department of Physics, Hubei University, Wuhan 430062, China}
\author{Wan-Li Yang}
\email{ywl@wipm.ac.cn}
\affiliation{State Key Laboratory of Magnetic Resonance and Atomic and Molecular Physics, Innovation Academy for Precision Measurement Science and Technology, Chinese Academy of Sciences, Wuhan 430071, China}
\author{Jun-Hong An\orcid{0000-0002-3475-0729}}
\email{anjhong@lzu.edu.cn}
\affiliation{Key Laboratory of Quantum Theory and Applications of MoE, Lanzhou Center for Theoretical Physics, and Key Laboratory of Theoretical Physics of Gansu Province, Lanzhou University, Lanzhou 730000, China}

\begin{abstract}
The quantum battery (QB) makes use of quantum effects to store and supply energy, which may outperform its classical counterpart. However, there are two challenges in this field. One is that the environment-induced decoherence causes the energy loss and aging of the QB, the other is that the decreasing of the charger-QB coupling strength with increasing their distance makes the charging of the QB become inefficient. Here, we propose a QB scheme to realize a remote charging via coupling the QB and the charger to a rectangular hollow metal waveguide. It is found that an ideal charging is realized as long as two bound states are formed in the energy spectrum of the total system consisting of the QB, the charger, and the electromagnetic environment in the waveguide. Using the constructive role of the decoherence, our QB is immune to the aging. Additionally, without resorting to the direct charger-QB interaction, our scheme works in a way of long-range and wireless-like charging. Effectively overcoming the two challenges, our result supplies an insightful guideline to  the practical realization of the QB by reservoir engineering.
\end{abstract}

\maketitle

\section{Introduction} The miniaturization of electronic devices has inspired a rapid development of quantum technologies, which are expected to bring a lot of technological innovations \cite{RevModPhys.94.015004,RevModPhys.92.025002,RevModPhys.89.035002}. Among them, quantum thermodynamics \cite{Mayer2023,Maslennikov2019,Alicki2018} has emerged as a field aiming to reconstruct thermodynamics via basic laws of quantum mechanics and to break through the constraint of classical physics to device performance by quantum effects \cite{PhysRevLett.122.110601,Campaioli2018,PhysRevLett.112.030602,PhysRevE.87.042123,PhysRevLett.105.130401,Scully2003}. As an energy-storing and -converting device in atomic size, the quantum battery (QB) promotes the application of quantum effects in thermodynamics \cite{campaioli2023colloquium,Bhattacharjee2021}. With the aid of quantum resources, such as quantum coherence and entanglement, the QB possesses a potential of stronger charging power, higher charging capacity, and larger work extraction than its classical counterpart \cite{PhysRevLett.131.030402,PhysRevLett.130.210401,PhysRevLett.129.130602,PhysRevLett.125.180603,PhysRevE.102.042111,RevModPhys.91.025001,PhysRevLett.111.240401}.

An efficient QB scheme is a prerequisite for its realization and further applications. The widely studied QB model is based on two-level systems (TLSs), which are charged by the coherent coupling to other TLSs \cite{PhysRevResearch.5.013155,PhysRevE.106.054119,PhysRevLett.127.100601,PhysRevB.104.245418,PhysRevB.98.205423} or external fields \cite{PhysRevA.107.032203,PhysRevA.107.012207,PhysRevResearch.4.L022017,PhysRevA.106.062609,PhysRevResearch.3.L032073,PhysRevResearch.2.023095,PhysRevE.100.032107}. Experiments on the single-body QB were carried out in nuclear magnetic resonance \cite{PhysRevA.106.042601}, superconducting circuit \cite{Zheng_2022,Hu_2022}, semiconducting quantum dot \cite{PhysRevLett.131.260401,batteries8050043}, and photonic \cite{PhysRevLett.131.240401} systems. It was also found that the collection of more TLSs exhibits more energy storage capacity and the collective charging benefiting from quantum correlation induces stronger charging power of the QB \cite{PhysRevA.107.L030201,ref1,PhysRevResearch.4.013172,PhysRevB.105.115405,PhysRevA.106.022618,PhysRevA.105.022628,PhysRevA.106.032212,PhysRevLett.128.140501,PhysRevE.105.054115,Barra_2022,PhysRevA.103.033715,PhysRevE.104.024129,PhysRevA.101.032115,PhysRevE.102.052109,PhysRevLett.125.236402,Rosa2020,PhysRevB.100.115142,PhysRevB.99.205437,PhysRevA.97.022106,PhysRevLett.120.117702,PhysRevLett.118.150601,Binder_2015}. However, the practical performance of the QB is challenged by two facts. One is that its energy during the long-time storage process is depleted by the environment-induced decoherence, which is called the aging of the QB \cite{PhysRevA.100.043833}. The other is that its charging becomes inefficient when the charger-QB coupling is weakened due to certain reasons. Several schemes using feedback control \cite{PhysRevE.106.014138}, dark states \cite{PhysRevApplied.14.024092}, Floquet engineering \cite{PhysRevA.102.060201}, and environment engineering \cite{PhysRevA.106.012425,PhysRevE.105.064119,PhysRevA.105.062203,PhysRevE.104.064143,PhysRevE.104.064143,PhysRevA.104.032207,PhysRevE.104.044116,PhysRevLett.122.210601,PhysRevB.99.035421,PhysRevE.106.054107,PhysRevA.104.032606,PhysRevA.102.052223,Kamin_2020}, have been proposed to beat the decoherence effects. However, how to suppress the decoherence effects on the QB and improve the charging efficiency simultaneously is still an open question. 

\begin{figure}
    \includegraphics[width=\columnwidth]{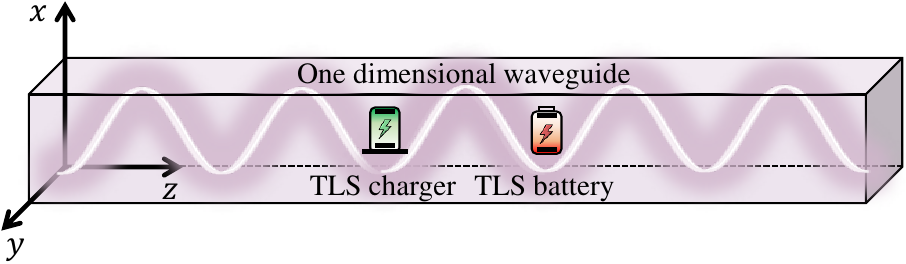}
    \caption{Scheme of a remote-charging QB. The noninteracting charger and QB formed by TLSs are separated in a distance $\Delta_z$ in a rectangular hollow metal waveguide. An efficient remote charging to the QB is realized by the mediation role of the EMF in the waveguide.}
    \label{fig:model}
\end{figure}

Inspired by the development of quantum interconnect to establish correlation among distant quantum entities \cite{PRXQuantum.2.017002}, we propose a remote-charging QB scheme via coupling the QB and the charger to a rectangular hollow metal waveguide \cite{PhysRevLett.131.073602,PhysRevLett.127.083602,Song:17}. Utilizing the non-Markovian decoherence induced by the electromagnetic field (EMF) in the waveguide, we find a mechanism that allows for a persistent energy exchange between the noninteracting charger and QB in the steady state. It is due to the formation of two bound states in the energy spectrum of the total system formed by the charger, the QB, and the EMF. Not causing the aging of the QB, the decoherence in our scheme plays a constructive role in the charging. Without resorting to the direct charger-QB coupling, our charging works in a long-range and wireless way. Overcoming the aging and low charging efficiency in the weak charger-QB coupling condition simultaneously, our scheme supplies an insightful guideline to the practical realization of the QB.

\section{Remote-charging QB model} We consider a long-range charging QB scheme, which consists of two TLSs acting as the charger and the QB residing in a rectangular hollow metal waveguide (see Fig. \ref{fig:model}). Without direct interaction, the charger and QB couple to a common EMF in the waveguide. The Hamiltonian reads \cite{PhysRevA.87.033831,Song:17,PhysRevLett.127.083602}
\begin{eqnarray}  \hat{H}&=&\omega_0\sum_{j=1,2}\hat{\sigma}^\dag_j\hat{\sigma}_j+\sum^{+\infty}_{k=-\infty}\big[\omega_k\hat{a}^\dagger_k\hat{a}_k\nonumber\\    &&+\sum_{j=1,2}\big(g_{k,j}\hat{\sigma}^\dag_j\hat{a}_k+\text{H.c.}\big)\big],\label{hamtl}
\end{eqnarray}
where $\hat{\sigma}_j=|g_j\rangle\langle e_j|$ is the transition operator from the excited state $|e_j\rangle$ to the ground state $| g_j \rangle$ with frequency $\omega_0$ of the charger when $j=1$ and of the QB when $j=2$ and $\hat{a}_k$ is the annihilation operator of the $k$th mode with frequency $\omega_k$ of the EMF. The coupling strength between the $j$th TLS and the $k$th mode of the EMF is $g_{k,j}=\sqrt{\omega_k/(2\epsilon_0)}\mathbf{d}_j\cdot\mathbf{u}_k(\mathbf{r}_j)$, where $\epsilon_0$ is the vacuum permittivity, $\mathbf{d}_j$ is the dipole moment of the $j$th TLS, and $\mathbf{u}_k(\mathbf{r}_j)$ is the spatial function of the EMF. The dispersion relation of the EMF is $\omega_k=[(ck)^2+\omega^2_{mn}]^{1/2}$, where $c$ is the speed of light, $k$ is the longitudinal wave number, and $\omega_{mn}=c[(m\pi/a)^2+(n\pi/b)^2]^{1/2}$, with $a$ and $b$ being the transverse lengths of the waveguide, is the cutoff frequency of the transverse mode of the EMF.
Assuming that the two TLSs are polarized in the $z$ direction, i.e., $\mathbf{d}_1=\mathbf{d}_2=d_z\mathbf{z}$, and considering only the dominated EMF mode with $m=n=1$, the coupling strength is further characterized by the spectral density
\begin{equation}    
J_{|j-j'|}(\omega)=\frac{\Gamma_{11}}{2\pi}\frac{\cos[\frac{z_j-z_{j'}}{c}\sqrt{\omega^2-\omega^2_{11}}]}{\sqrt{(\omega/\omega_{11})^2-1}}\Theta(\omega-\omega_{11}). 
\end{equation}
Here, $\Gamma_{11}=\frac{4\omega_{11}d_{z}^{2}}{\epsilon _{0} abc}\sin^{2}\left(\pi x_{0}/a\right)\sin^{2}\left(\pi y_0/b\right)$ is the radiation rate of the TLS with transverse position $\left(x_0,y_0\right)$ into the waveguide and $\Theta(\omega-\omega_{11})$ is the Heaviside step function. The spectral density $J_0(\omega)$ induces an individual spontaneous emission and Lamb frequency shift to the $j$th TLS, while $J_1(\omega)$ induces a correlated spontaneous emission and dipole-dipole interaction between the two TLSs. Different from Refs. \cite{PhysRevLett.122.210601,PhysRevLett.120.117702,PhysRevLett.127.100601,PhysRevLett.128.140501,PhysRevLett.125.236402}, we use such an incoherent correlation to charge the QB.  

Our scheme begins from a pump laser to initialize the charger to its excited state. When a charging to the QB is needed, the QB is taken to the waveguide not necessarily too near to the charger. The EMF in the waveguide is expected to mediate a charger-QB energy transfer. This process under the condition $|\Psi(0)\rangle=|e_1,g_2;\{0_k\}\rangle$ is described by $|\Psi(t)\rangle=[\sum_jc_j(t)\hat{\sigma}_j^\dag+\sum_k d_k(t)\hat{a}_k^\dag]| g_1,g_2,\{0_k\}\rangle$. Here, $c_j(t)$ and $d_k(t)$ are the excited-state probability amplitudes of the TLSs and the EMF with only one photon in the $k$th mode, respectively. From the Schr{\"o}dinger equation, we obtain \cite{SupplementalMaterial}
\begin{equation}
    \dot{c}_j(t)+i\omega_0 c_j(t)+\sum_{j'=1,2}\int^t_0 f_{|j-j'|}(t-\tau)c_{j'}(\tau)d\tau=0,
    \label{eq:dynamics}
\end{equation}
where $f_{|j-j'|}(x)=\int^{\infty}_{0}J_{|j-j'|}(\omega)e^{-i\omega x}d\omega$, $c_1(0)=1$, and $c_2(0)=0$. Equation \eqref{eq:dynamics} reveals that, although the direct charger-QB interaction is absent, an effective charger-QB coupling is induced by the mediation role of the EMF in the waveguide. The convolution in Eq. \eqref{eq:dynamics} renders the dynamics non-Markovian. The energy of the QB is 
\begin{equation}
    \mathcal{E}(t)=\text{Tr}[\rho_B(t)\hat{H}_B],
\end{equation}
where $\hat{H}_B=\omega_0\hat{\sigma}_2^\dagger\hat{\sigma}_2$ and $\rho_B(t)=\text{Tr}_1[|\Psi(t)\rangle\langle \Psi(t)|]$. According to the second law of thermodynamics, not all of $\mathcal{E}(t)$ can be converted into work \cite{PhysRevLett.124.130601}. The maximum of the extractable energy called ergotropy is defined as \cite{PhysRevLett.125.180603,PhysRevE.102.042111}
\begin{eqnarray}
    \mathcal{W}(t)=\text{Tr}[\rho_B(t)\hat{H}_B]-\text{Tr}[\tilde\rho_B(t)\hat{H}_B] ,
\end{eqnarray}
where $\tilde\rho_B(t)=\sum_k{r_k(t) |\varepsilon_k\rangle\langle\varepsilon_k|}$ is the passive state, $r_k(t)$ are the eigenvalues of $\rho_B(t)$ ordered in a descending sort, and $|\varepsilon_k\rangle$ are the eigenstates of $\hat{H}_B$ with the corresponding eigenvalues $\varepsilon_k$ ordered in an ascending sort.

In the special case of the weak TLS-EMF coupling, we apply the Markovian approximation to neglect the memory effect and extend the upper limit of the $\tau$ integral in Eq. \eqref{eq:dynamics} to infinity. The approximate solution is
\begin{eqnarray}
c_{j}^\text{MA}(t)=e^{-(i\omega_0+\Upsilon_0) t}[e^{-\Upsilon_1 t}-(-1)^j e^{\Upsilon_1 t}]/2,\label{cat}
\end{eqnarray}where $\Upsilon_j=\pi J_j(\omega_0)+i\delta_j$, with $\delta_j=\mathcal{P}\int_0^\infty d\omega{J_j(\omega)\over\omega_0-\omega}$ and $\mathcal{P}$ denoting the Cauchy principal value. Equation \eqref{cat} reveals that the EMF in the waveguide indeed can mediate an energy exchange between the charger and the QB. However, it is transient and the energies in both of them exponentially damp to zero in the long-time condition due to the destructive role of the Markovian decoherence. 

In contrast to the above approximate result, the mediated charger-QB coupling can induce a persistently reversible energy exchange between them even in the steady state in the non-Markovian dynamics. Equation \eqref{eq:dynamics} is only numerically solvable. However, its long-time form is analytically obtainable by the Laplace transform method. It converts Eq. \eqref{eq:dynamics} to
$\tilde{c}_{j}(s)=[{1\over \Xi(s)+\tilde{f}_1(s)}-{(-1)^j\over \Xi(s)-\tilde{f}_1(s)} ]/2$, where $\Xi(s)=s+i\omega_0+\tilde{f}_0(s)$ and $\tilde{f}_{|j-j'|}(s)=\int_0^{\infty}\frac{J_{|j-j'|}(\omega)}{i\omega+s}d\omega$. $c_j(t)$ is obtained by making the inverse Laplace transform to $\tilde{c}_j(s)$, which needs finding the poles of $\tilde{c}_j(s)$ from 
\begin{equation}
     Y_\pm(E)\equiv\omega_0-\int^\infty_{0}\frac{J_0(\omega)\pm J_1(\omega)}{\omega-E}d\omega=E,(E=is).     \label{eq:ES}
\end{equation}
Note that the roots $E$ of Eq. \eqref{eq:ES} are just the eigenenergies of Eq. \eqref{hamtl}. To prove this, we expand the eigenstate as $|\Phi\rangle=[\sum_j \alpha_j\hat{\sigma}_j^\dag+\sum_k \beta_k\hat{a}_k^\dag]|g_1,g_2,\{0_k\}\rangle$. From $\hat{H}|\Phi\rangle=E|\Phi\rangle$, we obtain $(E-\omega_0)\alpha_j=\sum_{j'}\int_{0}^\infty{J_{|j-j'|}(\omega)\alpha_{j'}\over E-\omega}d\omega$, which readily results in Eq. \eqref{eq:ES} after eliminating $\alpha_j$ \cite{SupplementalMaterial}. Thus, on one hand, Eq. \eqref{eq:ES} governs the dynamical evolution of $c_j(t)$, and on the other hand, it determines the eigenenergies of the total system. This means that the dynamics of the TLSs is essentially determined by the energy-spectrum feature of the total system. $Y_\pm(E)$ is ill-defined and jumps rapidly between $\pm\infty$ in the regime $E>\omega_{11}$ because of the divergence of the integrand. Thus, Eq. \eqref{eq:ES} has infinite roots in $E>\omega_{11}$, which form a continuous energy band. In the regime $E <\omega_{11}$, $Y_\pm(E)$ are monotonic decreasing functions of $E$ and an isolated root $E_+^\text{b}$ or $E_-^\text{b}$ is formed provided $Y_+(\omega_{11})<\omega_{11}$ or $Y_-(\omega_{11})<\omega_{11}$. We call the eigenstates corresponding to $E_\pm^\text{b}$ bound states. Substituting these poles from Eq. \eqref{eq:ES} into the inverse Laplace transform, the bound-state energies $E^\text{b}_\pm$ contribute two nontrivial residues; while the band energies in the continuum contribute a branch cut, which tends to zero in the long-time limit due to the out-of-phase interference \cite{PhysRevA.93.020105,PhysRevB.106.115427}. Thus, we have 
\begin{equation}
c_{j}(\infty)=\begin{cases}0, ~~~~~~~~~~~~~~~~~~~~~~~~~~~~~~~~~~~\text{0 BS},\\ 
Z_{+}e^{-iE^\text{b}_+t}, ~~~~~~~~~~~~~~~~~~~~~~~~\text{1 BS},\\Z_{+}e^{-iE^\text{b}_+t}-(-1)^jZ_{-}e^{-iE^\text{b}_-t}, \text{2 BSs},\end{cases}\label{ltmcl}
\end{equation}
where $Z_{\pm}={1\over 2}\big[1+\int^\infty_0\frac{J_0(\omega)\pm J_1(\omega)}{(E_\pm^\text{b}-\omega)^2}d\omega\big]^{-1}$ is the residue contributed by $E^\text{b}_\pm$ to $c_{j}(t)$. Equation \eqref{ltmcl} indicates the decisive role played by the energy-spectrum feature of the total TLS-EMF system in the non-Markovian dynamics of the TLSs. It is remarkable to see that the formation of the bound states prevents $|c_j(t)|^2$ from damping to zero, which is not captured by the Markovian result in Eq. \eqref{cat} and can be used to remotely charge the QB. When the QB is not needed for supplying energy to other devices, it is kept in the waveguide. Reversibly exchanging energy with the charger, the QB does not experience the aging caused by the energy loss in this storage process \cite{SupplementalMaterial}.

Equation \eqref{ltmcl} reveals that the QB is dynamically synchronous with the charger in the steady state. In the absence of the bound state, no energy is left in both the QB and the charger and the energy completely relaxes into the EMF of the waveguide.  If one bound state with eigenenergy $E^\text{b}_+$ is formed, then a stable energy $\mathcal{E}(\infty)=\omega_0 Z_+^2$ is preserved in the QB, which is equal to the one in the charger. As long as two bound states with eigenenergies $E^\text{b}_\pm$ are formed, a periodic energy exchange of the QB with the charger in a frequency $E^\text{b}_+-E^\text{b}_-$, i.e., $\mathcal{E}(\infty)=\omega_0\{Z^2_++Z^2_--2Z_+Z_-\cos[(E_+^\text{b}-E_-^\text{b})t]\}$, is kept in the steady state. Such a Rabi-like oscillation is naturally expected in the case of the direct charger-QB coupling. However, the charging realized by the direct coupling is generally fragile to the environment-induced decoherence, which results in the aging of the QB. Our result indicates that, different from the destructive role in the direct-coupling case, the non-Markovian decoherence caused by the common EMF can be used to realize an ideal charging to the QB in the open system framework. It is just the constructive role played by the decoherence that generates such an efficient charging in our scheme. 

The persistent Rabi-like charger-QB energy exchange endows our scheme with a substantial difference from previous environment-assisted charging schemes either in the Markovian approximation \cite{PhysRevA.106.012425,PhysRevE.105.064119,PhysRevA.105.062203,PhysRevE.104.064143,PhysRevE.104.064143,PhysRevA.104.032207,PhysRevE.104.044116,PhysRevLett.122.210601,PhysRevB.99.035421} or in the non-Markovian dynamics \cite{PhysRevE.106.054107,PhysRevA.104.032606,PhysRevA.102.052223,Kamin_2020}. Our charging performance, which is as perfect as the ideal charging case by direct charge-QB interaction, is guaranteed by the feature of the energy spectrum of the total system consisting of the charger, the QB, and the EMF in the waveguide. It makes our scheme immune to the environment-induced aging. Our scheme works in a way of long-range and wireless-like charging, while the spatial charger-QB distance has never been investigated. As an attractive feature in future applications, it overcomes the low-charging-efficiency problem experienced by the conventional QB schemes when the charger-QB coupling is weak.

\begin{figure}[t]
    \includegraphics[width=\columnwidth]{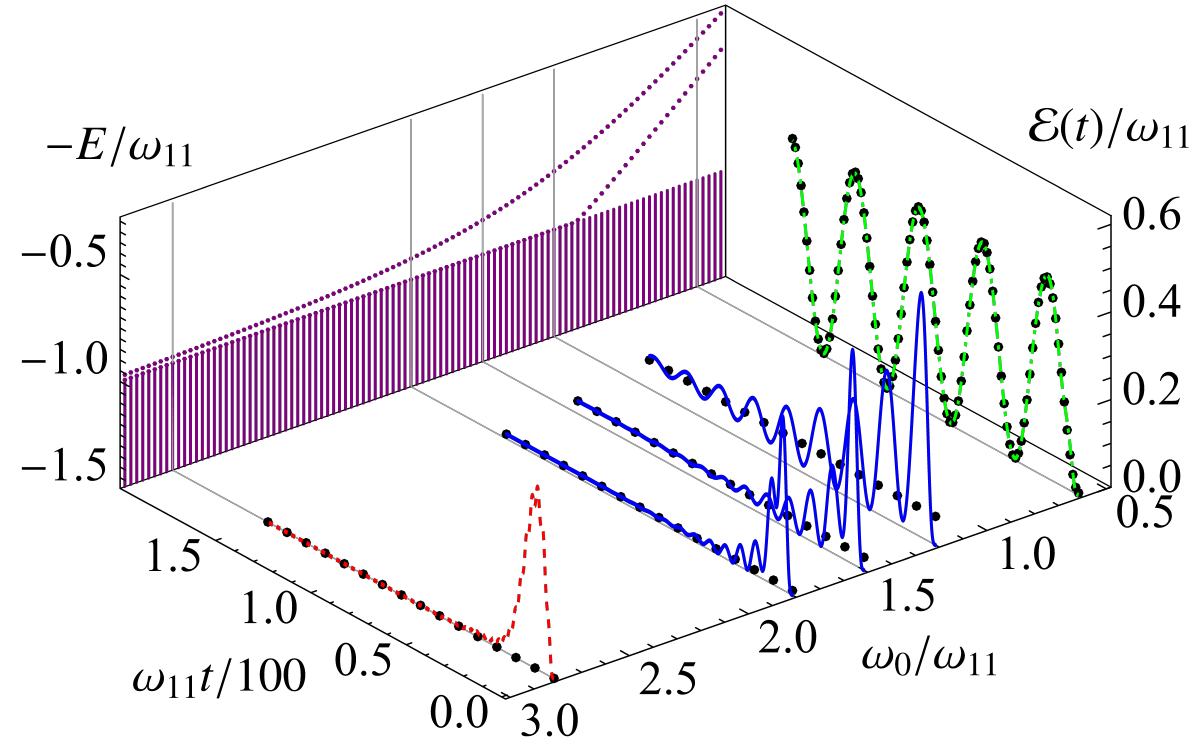}
    \caption{Energy spectrum $E$ of the total system via solving Eq. \eqref{eq:ES} and evolution of the QB energy $\mathcal{E}(t)$ via solving Eq. \eqref{eq:dynamics} in different $\omega_0$ in the presence of zero (red dashed lines), one (blue solid lines), and two (green dot-dashed lines) bound states. The black dots mark the results evaluated from the analytical equation \eqref{ltmcl}. We use $\Delta_z=0.1\lambda_{11}$ and $\Gamma_{11}=0.5\omega_{11}$.}
    \label{fig:Delta_z}
\end{figure}
\begin{figure}	\includegraphics[width=\columnwidth]{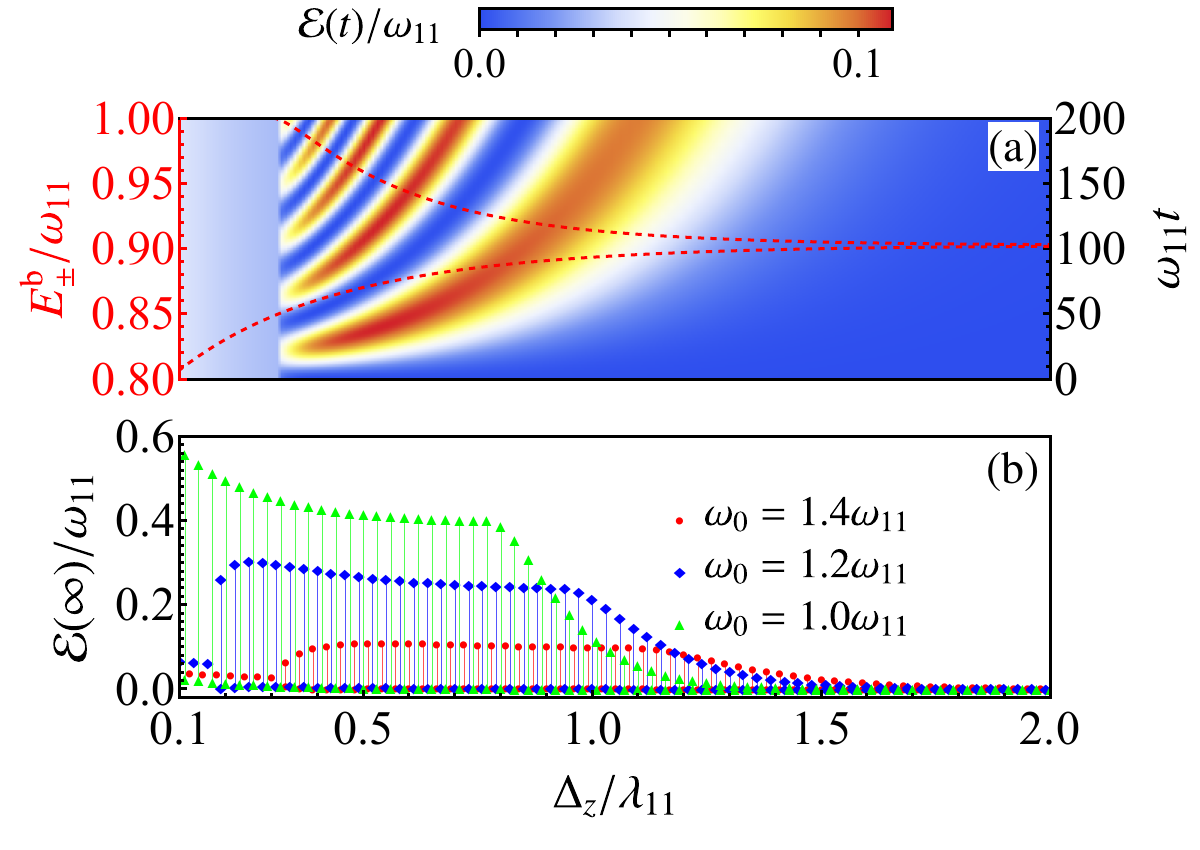}
	\caption{(a) Bound-state eigenenergies $E_\pm^\text{b}$ (red dashed lines) and long-time evolution of $\mathcal{E}(t)$, and (b) extremal values of $\mathcal{E}(\infty)$ as a function of $\Delta_{z}$ when $\omega_0=1.4\omega_{11}$. The cases of $\omega_0=1.2\omega_{11}$ (blue blocks) and $1.0\omega_{11}$ (green triangles) are also given in (b). The other parameters are the same as Fig. \ref{fig:Delta_z}.}	\label{fig:SED}
\end{figure}
\section{Numerical results}
Choosing the charger-QB distance $\Delta_z\equiv|z_1-z_2|=0.1\lambda_{11}$, with $\lambda_{11}=2\pi c/\omega_{11}$, which is much larger than the typical range of the dipole-dipole interactions, we plot in Fig. \ref{fig:Delta_z} the energy spectrum of the total system formed by the QB, the charger, and the EMF and the evolution of the QB energy $\mathcal{E}(t)$ in different frequency $\omega_0$ of the TLSs. It is interesting to find that, in contrast to the damping to zero in the Markovian case, $\mathcal{E}(t)$ approaches a finite value for a moderate $\omega_0$, while it exhibits a lossless Rabi-like oscillation for a small $\omega_0$. These diverse behaviors can be explained by the energy-spectrum feature. We see that two branches of bound states in the band gap region separate the energy spectrum into three regimes. When $\omega_0\ge2.8\omega_{11}$, no bound state is formed and $\mathcal{E}(t)$ damps to zero exclusively. When $1.1\omega_{11}\le\omega_0<2.8\omega_{11}$, one bound state is formed and $\mathcal{E}(t)$ tends to a finite value. As long as two bound states are present when $\omega_0<1.1\omega_{11}$, $\mathcal{E}(t)$ exhibits a lossless Rabi-like oscillation. The matching of the long-time behaviors in the three regimes with the analytical result in Eq. \eqref{ltmcl} verifies the distinguished role played by the bound states and non-Markovian effect in the charging performance of our scheme. The result reveals that we can properly choose the working frequency $\omega_0$ of the QB and charger such that a remote charging to the QB is dynamically realized without resorting to the direct charger-QB interactions.

\begin{figure}	\includegraphics[width=\columnwidth]{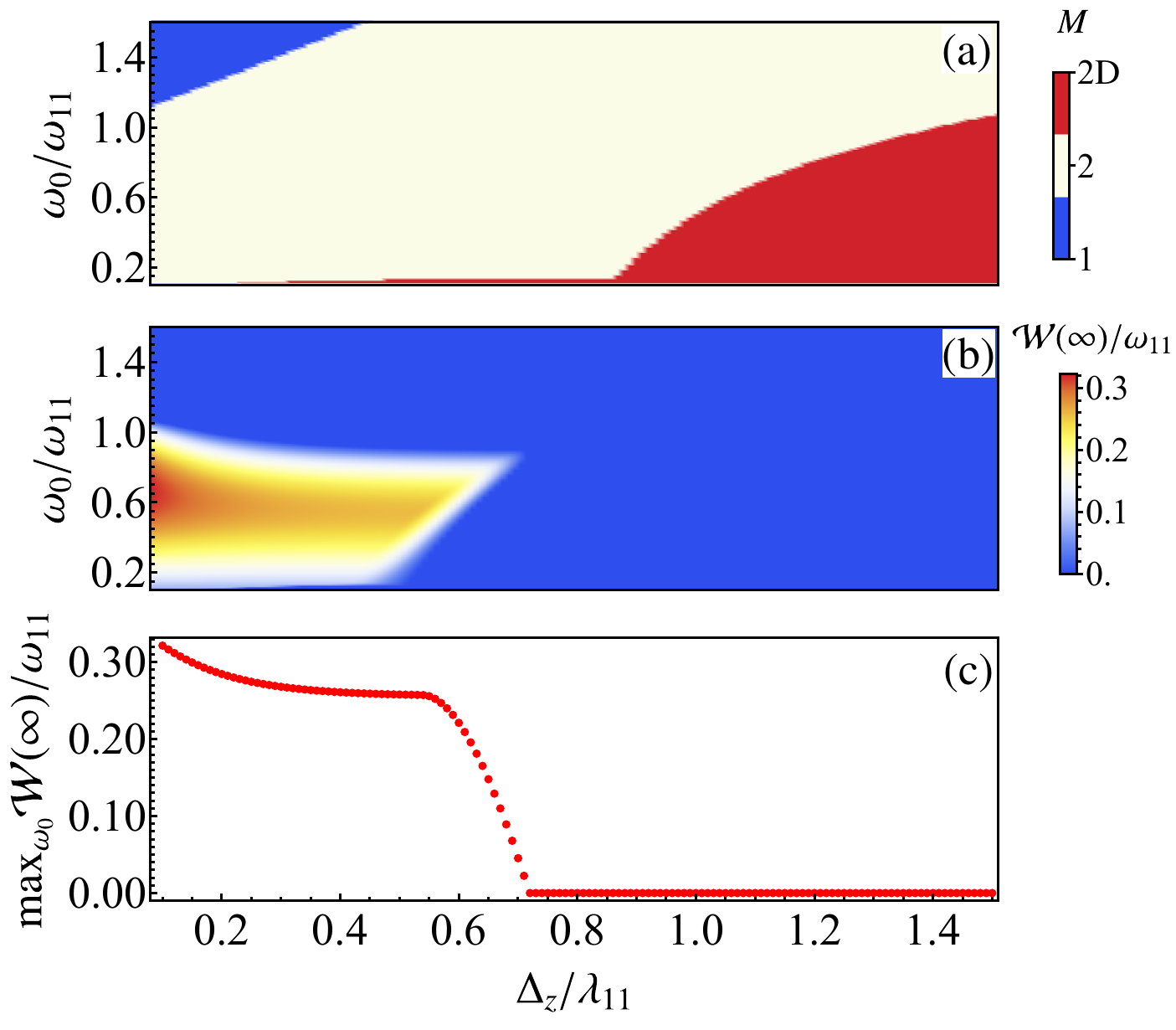}
\caption{(a) Number $M$ of the bound states, (b) maxima of the ergotropy $\mathcal{W}(\infty)$ in different $\omega_0$, and (c) maxima of the ergotropy $\text{max}_{\omega_0}\mathcal{W}(\infty)$ via optimizing $\omega_0$ as a function of $\Delta_z$. 2D means two degenerate bound states. The other parameters are the same as Fig. \ref{fig:Delta_z}.}	\label{fig:EGD}
\end{figure}

To verify the tolerance of the scheme to the increase of the charger-QB distance, we plot in Fig. \ref{fig:SED} the eigenenergies of bound states, the long-time evolution of $\mathcal{E}(t)$, and the extremal values of $\mathcal{E}(\infty)$ as a function of $\Delta_z$. The result for $\omega_0=1.4\omega_{11}$ shows that when $\Delta_z<0.3\lambda_{11}$, one bound state is present and $\mathcal{E}(t)$ evolves to a finite $\Delta_z$-dependent value. When $0.3\lambda_{11}\le\Delta_z<1.7\lambda_{11}$, two bound states are formed and $\mathcal{E}(t)$ tends to a persistent Rabi-like oscillation in a frequency $E_+^\text{b}-E_-^\text{b}$ characterizing a coherent energy exchange between the QB and the charger. When $\Delta_z\ge 1.7\lambda_{11}$, the eigenenergies of the two bound states become degenerate and $\mathcal{E}(t)$ tends to zero according to Eq. \eqref{ltmcl}. Therefore, our charging scheme can work well when $\Delta_z$ is freely changed within the range from $0.3\lambda_{11}$ to $1.7\lambda_{11}$. The result in Fig. \ref{fig:SED}(b) also reveals that the maximal stored energy $\max\mathcal{E}(\infty)$ can be enhanced by properly choosing the working frequency $\omega_0$ of the QB and the charger. This gives us sufficient room to optimize the performance of our remote-charging scheme. Note that the presence of two bound states in Fig. \ref{fig:EGD}(a) and nonzero $\mathcal{E}(\infty)$ are not always accompanied by an extractable energy of the QB described by the nonzero ergotropy $\mathcal{W}(\infty)$, see Fig. \ref{fig:EGD}(b). Via optimizing $\omega_0$, we plot $\max_{\omega_0}\mathcal{W}(\infty)$ in Fig. \ref{fig:EGD}(c) in different $\Delta_z$, which exhibits a good performance of our QB in quite a large charger-QB distance regime. 

\section{Discussions and conclusion} Although only the rectangular hollow metal waveguide is considered, our result is applicable to other systems. Applying it in the surface plasmon polariton waveguide \cite{Falk2007,PhysRevLett.106.020501}, we are even not bothered to put the charger and the QB inside the waveguide, which reduces the experimental difficulty. Current experimental advances of the waveguide QED provide support for our QB scheme \cite{RevModPhys.95.015002,PhysRevLett.122.173603,Mirhosseini2019,science.aah6875}. Especially, based on the solid-state emitter, i.e., nitrogen-vacancy centers (NVC), a series of integrated platforms have ensured the feasibility of our proposal, such as microwave coplanar \cite{PhysRevLett.107.060502}, silicon nitride \cite{PhysRevX.5.031009}, laser-written \cite{PhysRevApplied.15.054059}, and nanowire \cite{PhysRevApplied.17.014008} waveguides. The long-range interaction of NVCs in magnon waveguide QED has been proposed \cite{PRXQuantum.2.040314}. In our setup, proper values of the working frequency $\omega_0$ and the charger-QB distance $\Delta_z$ are required. NVC is of the capability for coherent manipulation by either light or microwave. In the optical frequency regime, we can focus on the working frequency $\lambda_0=637$ nm. To match $\omega_0=1.0\omega_{11}$ for the maximal stored energy $\mathcal{E}=0.6\omega_{11}$, the transverse lengths of waveguide are set as $a=b=0.45$ $\mu$m. To match $\omega_0=0.6\omega_{11}$ for the maximal extractable energy $\mathcal{W}=0.3\omega_{11}$, they are set as $a=b=0.27$ $\mu$m. In the microwave frequency regime, the working frequency of NVC with zero-field splitting $2.87$ GHz has a tremendous tunability by an external field. Generally speaking, the effective charging distance could range from $1$ $\mu$m in the optical waveguide to $0.1$ m in the microwave one, which are far beyond the typical length of the dipole-dipole coupling. The bound state and its distinguished role in the non-Markovian dynamics have been observed in both photonic crystal \cite{Liu2017} and ultracold-atom \cite{Krinner2018,Kwon2022} systems, which sets a solid foundation to the realizability of our scheme. Also, presenting a scalable and tunable framework, our scheme is easily extended to the multi-TLS scenarios.

In conclusion, we have proposed a remote-charging scheme via coupling the QB and the charger to a rectangular hollow metal waveguide. It is found that, contrary to one's belief that decoherence generally leads to the aging of a QB, the non-Markovian decoherence induced by the EMF in the waveguide can realize a persistent Rabi-like energy exchange between the QB and the charger. Our analysis reveals that such an ideal charging performance is caused by the formation of two bound states in the energy spectrum of the total system consisting of the QB, the charger, and the EMF in the waveguide. Without resorting to direct charger-QB coupling, our scheme avoids the insufficient-charging difficulty encountered in the case when the charger-QB coupling is weak. It provides a useful path to the practical realization of antiaging QB.

This work is supported by the National Natural Science Foundation of China (Grants No. 12105089,  No. 12074107, No. 12275109, No. 12274422, and No. 12247101), the innovation group project (Grant No. 2022CFA012) of Hubei Province, the program of outstanding young and middle-aged scientific and technological innovation team of colleges and universities in Hubei Province (Grant No. T2020001), and the Hubei Province Science Fund for Distinguished Young Scholars under Grant No. 2020CFA078. 

\bibliography{QB}

\end{document}


\title{Supplemental Material for ``Remote Charging and Degradation Suppression for the Quantum Battery''}
\author{Wan-Lu Song\orcid{0000-0002-6437-9748}}
\affiliation{Department of Physics, Hubei University, Wuhan 430062, China}
\author{Hai-Bin Liu}
\affiliation{Department of Physics, Hubei University, Wuhan 430062, China}
\author{Bin Zhou\orcid{0000-0002-4808-0439}}
\email{binzhou@hubu.edu.cn}
\affiliation{Department of Physics, Hubei University, Wuhan 430062, China}
\author{Wan-Li Yang}
\email{ywl@wipm.ac.cn}
\affiliation{State Key Laboratory of Magnetic Resonance and Atomic and Molecular Physics, Innovation Academy for Precision Measurement Science and Technology, Chinese Academy of Sciences, Wuhan 430071, China}
\author{Jun-Hong An\orcid{0000-0002-3475-0729}}
\email{anjhong@lzu.edu.cn}
\affiliation{Key Laboratory of Quantum Theory and Applications of MoE, Lanzhou Center for Theoretical Physics, and Key Laboratory of Theoretical Physics of Gansu Province, Lanzhou University, Lanzhou 730000, China}
\maketitle

We consider two two-level systems, which act as a charger and a quantum battery (QB), residing in a rectangular hollow metal waveguide. They couple to a common electromagnetic field (EMF) in the waveguide. The Hamiltonian of the total system is \cite{PhysRevA.87.033831,Song:17,PhysRevLett.127.083602}
\begin{eqnarray}  
\hat{H}&=&\omega_0\sum_{j=1,2}\hat{\sigma}^\dag_j\hat{\sigma}_j+\sum^{+\infty}_{k=-\infty}\big[\omega_k\hat{a}^\dagger_k\hat{a}_k\nonumber\\    &&+\sum_{j=1,2}\big(g_{k,j}\hat{\sigma}^\dag_j\hat{a}_k+\text{H.c.}\big)\big],\label{hamtl}
\end{eqnarray}
where $\hat{\sigma}_j=|g_j\rangle\langle e_j|$ is the transition operator from the excited state $|e_j\rangle$ to the ground state $| g_j \rangle$ with frequency $\omega_0$ of the charger when $j=1$ and of the QB when $j=2$ and $\hat{a}_k$ is the annihilation operator of the $k$th mode with frequency $\omega_k$ of the EMF. The coupling strength between the $j$th TLS and the $k$th mode of the EMF is $g_{k,j}=\sqrt{\omega_k/(2\epsilon_0)}\mathbf{d}_j\cdot\mathbf{u}_k(\mathbf{r}_j)$, where $\epsilon_0$ is the vacuum permittivity, $\mathbf{d}_j$ is the dipole moment of the $j$th TLS, and $\mathbf{u}_k(\mathbf{r}_j)$ is the spatial function of the EMF in the waveguide. The dispersion relation of the EMF in the waveguide is $\omega_k=[(ck)^2+\omega^2_{mn}]^{1/2}$, where $c$ is the speed of light, $k$ is the longitudinal wave number, and $\omega_{mn}=c[(m\pi/a)^2+(n\pi/b)^2]^{1/2}$, with $a$ and $b$ being the transverse lengths of the waveguide, is the cutoff frequency of transverse mode of the EMF. The corresponding spatial function $\mathbf{u}_k(\mathbf{r}_j)$ of the EMF includes the transverse electric (TE) and transverse magnetic (TM) modes labeled by $m$ and $n$ as
\begin{eqnarray}
\mathbf{u}_{mn,k}^\text{TE}(\mathbf{r}_j)&=&\frac{2c\pi e^{ikz_{j}}}{\sqrt{abL}}\Big[ -\frac{n}{b}\cos\frac{m\pi x_{j}}{a}\sin\frac{n\pi y_{j}}{b}\mathbf{e}_{x}\nonumber \\ 
&&+\frac{m }{a}\sin  \frac{m\pi x_{j} }{a} \sin  \frac{n\pi y_{j}}{b} \mathbf{e}_{y}\Big], \label{ute} \\
\mathbf{u}_{mn,k}^\text{TM}(\mathbf{r}_j)&=&\frac{2e^{ikz_{j}}}{\omega_k\sqrt{abL}}\{\frac{ikc^2\pi}{\omega_{mn}}\Big[\frac{m }{a}\cos\frac{m\pi x_{j}}{a} \sin\frac{n\pi y_{j}}{b}\mathbf{e}_{x}\nonumber\\
&&+\frac{n }{b}\sin \frac{m\pi x_{j}}{a} \sin \frac{n\pi y_{j}}{b} \mathbf{e}_{y}\Big] \nonumber\\
&&+\omega_{mn}\sin  \frac{m\pi x_{j}}{a}\sin \frac{n\pi y_{j}}{b} \mathbf{e}_{z}\}, \label{utm}
\end{eqnarray}
where $\mathbf{r}_j=(x_j,y_j,z_j)$ and $L$ is the length of the waveguide. We assume that the charger and the QB are polarized in the $z$ direction, i.e., $\mathbf{d}_1=\mathbf{d}_2=d_z\mathbf{z}$, and consider only the dominated EMF mode with $m=n=1$ in the waveguide. Then we obtain
\begin{equation}
g_{k,j}=e^{ikz_{j}}\omega_{11}d_{z}\sqrt{\frac{2}{\epsilon_{0}abL\omega_k}}\sin\frac{\pi x_{0}}{a}\sin\frac{\pi y_{0}}{b},\label{cst}
\end{equation}
where $(x_0, y_0)$ is the common transverse position of the charger and the QB. 

\section{Energy spectrum}
The total excitation number of the total system, i.e., $\hat{N}=\sum_j\hat{\sigma}_j^\dag\hat{\sigma}_j+\sum_k\hat{a}_k^\dag\hat{a}_k$, is conserved due to $[\hat{N},\hat{H}]=0$. Therefore, the eigenstate of the total system in the single-excitation subspace can be expanded as 
\begin{equation}
|\Phi\rangle=\Big[\sum_{j=1,2} \alpha_j\hat{\sigma}_j^\dag+\sum_{k=-\infty}^{+\infty} \beta_k\hat{a}_k^\dag\Big]|g_1,g_2,\{0_k\}\rangle.\label{eigst}
\end{equation} 
The substitution of Eq. \eqref{eigst} into $\hat{H}|\Phi\rangle=E|\Phi\rangle$ results in 
\begin{eqnarray}
E\alpha_j=\omega_0\alpha_j+\sum^{+\infty}_{k=-\infty} g_{k,j}\beta_k,\label{albe}\\
E\beta_k=\omega_k\beta_k+\sum_{j=1,2} g_{k,j}^*\alpha_{j}.\label{eabea}
\end{eqnarray}
Substituting the solution $\beta_k=\sum_j g_{k,j}^*\alpha_{j}/(E-\omega_k)$ of Eq. \eqref{eabea} into Eq. \eqref{albe}, we have
\begin{eqnarray}
\alpha_1\Big(E-\omega_0-\sum^{+\infty}_{k=-\infty} \frac{|g_{k,1}|^2}{E-\omega_k}\Big) =\alpha_2\sum^{+\infty}_{k=-\infty}\frac{g_{k,1}g^\ast_{k,2}}{E-\omega_k}, \label{ce1}\\
\alpha_2\Big(E-\omega_0-\sum^{+\infty}_{k=-\infty} |\frac{g_{k,2}|^2}{E-\omega_k}\Big) =\alpha_1\sum^{+\infty}_{k=-\infty}\frac{g_{k,2}g^\ast_{k,1}}{E-\omega_k}. \label{ce2}
\end{eqnarray}
We define the correlated spectral density $J_{jj'}(\omega)$ as
\begin{eqnarray}
J_{jj'}(\omega)=\sum_{k=-\infty}^{+\infty}  g_{k,j}g^\ast_{k,j'}\delta(\omega-\omega_k).\label{spctd}
\end{eqnarray}
Substituting Eq. \eqref{cst} into Eq. \eqref{spctd} and using $g_{-k,j}=g_{k,j}^*$, we obtain
\begin{equation}
J_{jj'}(\omega)=\frac{\Gamma_{11}}{2\pi}\frac{\cos\Big[\frac{z_j-z_{j'}}{c}\sqrt{\omega^2-\omega^2_{11}}\Big]}{\sqrt{(\omega/\omega_{11})^2-1}}\Theta(\omega-\omega_{11}), \label{sd}
\end{equation}
where $\Gamma_{11}=\frac{4\omega_{11}d_{z}^{2}}{\epsilon _{0}abc}\sin^{2}\frac{\pi x_{0}}{a}\sin^{2}\frac{\pi y_{0}}{b}$ and $\Theta(\omega-\omega_{11})$ is the Heaviside step function. Equation \eqref{sd} reveals $J_{jj'}(\omega)=J_{|j-j'|}(\omega)$. Thus, Eqs. (\ref{ce1}) and (\ref{ce2}) can be rewritten as
\begin{eqnarray}
\alpha_1\Big[E-\omega_0-\int_0^{\infty}\frac{J_0(\omega)}{E-\omega}d\omega\Big] =\alpha_2\int_0^{\infty}\frac{J_1(\omega)}{E-\omega}d\omega, ~\label{ce3}\\
\alpha_2\Big[E-\omega_0-\int_0^{\infty}\frac{J_0(\omega)}{E-\omega}d\omega\Big] =\alpha_1\int_0^{\infty}\frac{J_1(\omega)}{E-\omega}d\omega.~ \label{ce4}
\end{eqnarray}
Eliminating $\alpha_1$ and $\alpha_2$ from Eqs. (\ref{ce3}) and (\ref{ce4}), we obtain
\begin{equation}
\omega_0-\int^\infty_0\frac{J_0(\omega)\pm J_1(\omega)}{\omega-E}d\omega=E.\label{dds}
\end{equation}

\section{Dynamics}
Under the initial condition $|\Psi(0)\rangle=|e_1,g_2,\{0_k\}\rangle$, the time-evolved state is expanded as $|\Psi(t)\rangle=[\sum_jc_j(t)\hat{\sigma}_j^\dag+\sum_k d_k(t)\hat{a}_k^\dag]| g_1,g_2,\{0_k\}\rangle$, where $c_j(t)$ and $d_k(t)$ are the excited-state probability amplitudes of the TLSs and the EMF with only one photon in the $k$th mode, respectively. From the time-dependent Schr{\"o}dinger equation $i|\dot{\Psi}(t)\rangle=\hat{H}|\Psi(t)\rangle$, we obtain
\begin{eqnarray}
i\dot{c}_{1}(t)&=&\omega_{0}c_{1}(t)+\sum_{k=-\infty}^{+\infty}{g_{k,1}d_{k}(t)}, \label{c1t}\\
i\dot{c}_{2}(t)&=&\omega_{0}c_{2}(t)+\sum_{k=-\infty}^{+\infty}{g_{k,2}d_{k}(t)}, \label{c2t}\\
i\dot{d}_{k}(t)&=&\omega_{k}d_{k}(t)+\sum_{j=1,2}g_{k,j}^{\ast}c_{j}(t). \label{dkt}
\end{eqnarray}
From Eq. (\ref{dkt}), we have
\begin{equation}
d_{k}(t)=-i\sum_{j=1,2}\int_{0}^{t}e^{-i\omega_{k}(t-\tau)}g_{k,j}^{\ast}c_{j}(\tau)d\tau. \label{dk}
\end{equation}
The substitution of Eq. (\ref{dk}) into Eqs. (\ref{c1t}) and (\ref{c2t}) results in
\begin{equation}
\dot{c}_j(t)+i\omega_0 c_j(t)+\sum_{j'=1,2}\int^t_0 f_{|j-j'|}(t-\tau)c_{j'}(\tau)d\tau=0,\label{eq:dynamics}
\end{equation}
whose initial condition is $c_1(0)=1$ and $c_2(0)=0$. The kernel functions are $f_{|j-j'|}(t-\tau)=\int^{\infty}_{0}J_{|j-j'|}(\omega)e^{-i\omega(t-\tau)}d\omega$.

When the couplings between the two TLSs and the environment are weak, we can apply the Markovian approximation. Setting $c_j(t)=e^{-i\omega_0t}\underline{c}_j(t)$, Eq. \eqref{eq:dynamics} is recast into
\begin{equation}
\dot{\underline c}_j(t)+\sum_{j'=1,2}\int_0^td\tau {\underline c}_{j'}(\tau)\int_{0}^\infty d\omega e^{-i(\omega-\omega_0)(t-\tau)}J_{|j-j'|}(\omega)=0.\label{mastf}
\end{equation}
Under the Markovian approximation, we neglect the memory effect by replacing $\underline{c}_{j'}(\tau)$ with $\underline{c}_{j'}(t)$ and extend the upper limit of the $\tau$-integral from $t$ to infinity. Then, Eq. \eqref{mastf} becomes
\begin{equation}
\dot{\underline c}_j(t)+\sum_{j'=1,2}\Upsilon_{|j-j'|} {\underline c}_{j'}(t) =0,\label{mastf22}
\end{equation}
where $\Upsilon_{|j-j'|}=\pi J_{|j-j'|}(\omega_0)+i\delta_{|j-j'|}$, with $\delta_{|j-j'|}=\mathcal{P}\int_0^{\infty} d\omega \frac{J_{|j-j'|}(\omega)}{\omega_0-\omega}$ and $\mathcal{P}$ being the Cauchy principal value, and the identity $\int_0^\infty d\tau e^{-i(\omega-\omega_0)\tau}=\pi\delta(\omega-\omega_0)+i\mathcal{P}{1\over \omega_0-\omega}$ has been used. 
Solving the ordinary-differential equation \eqref{mastf22}, we have the Markovian approximate solution as
\begin{equation}
c_j^\text{MA}(t)=e^{-(i\omega_0+\Upsilon_0)t}[e^{-\Upsilon_1t}-(-1)^je^{\Upsilon_1t}]/2.\label{mkvr}
\end{equation}

In the non-Markovian dynamics, the exact solution of Eq. (\ref{mastf}) can only be obtained by numerical calculation. However, the long-time solution is analytically solvable by using the Laplace transform method. The Laplace transform $\Tilde{c}_j(s)=\int_0^{\infty} c_j(t)e^{-st}dt$ converts Eq. (\ref{mastf}) into
\begin{equation}
\tilde{c}_{j}(s)={1\over 2}\Big[{1\over \Xi(s)+\tilde{f}_1(s)}-{(-1)^j\over \Xi(s)-\tilde{f}_1(s)}\Big ],
\end{equation}
where $\Xi(s)=s+i\omega_0+\tilde{f}_0(s)$, with $\tilde{f}_{|j-j'|}(s)=\int_0^{\infty}\frac{J_{|j-j'|}(\omega)}{i\omega+s}d\omega$ being the Laplace transform of $f_{|j-j'|}(t)$. Then $c_j(t)$ can be obtained by applying the inverse Laplace transform to $\Tilde{c}_j(s)$, i.e., 
\begin{equation}
c_j(t)=\frac{1}{2\pi i}\int_{\sigma-i\infty}^{\sigma+i\infty}\Tilde{c}_j(s)e^{st}ds,\label{invlps}
\end{equation}
where $\sigma$ is chosen to be larger than all the poles of $\tilde{c}_j(s)$. The poles of  $\tilde{c}_j(s)$ are determined by
\begin{equation}
Y_\pm(E)=\omega_0-\int_0^{\infty}\frac{J_0(\omega)\pm J_1(\omega)}{\omega-E}d\omega=E,\label{evgpl}
\end{equation}
where $E=is$. It is interesting to see the the pole equation \eqref{evgpl} governing the dynamical feature of $c(t)$ is the same as the equation \eqref{dds} satisfied by the eigenenergy. It demonstrates that the dynamics of the open system is essentially determined by the feature of the energy spectrum of the total system. Due to the Heaviside step function $\Theta(\omega-\omega_{11})$ in Eq. \eqref{sd}, both $Y_+(E)$ and $Y_-(E)$ on the left side of Eq. \eqref{evgpl} are monotonically decreasing functions of $E$ in the regime $E<\omega_{11}$. Therefore, either the ``$+$'' branch or the ``$-$'' branch of Eq. \eqref{evgpl} has one and only one isolated root $E_+^\text{b}$ or $E_-^\text{b}$ in this regime provided $Y_+(\omega_{11})<\omega_{11}$ or $Y_-(\omega_{11})<\omega_{11}$, respectively. We call the corresponding eigenstates of these isolated roots $E_\pm^\text{b}$ the bound states. On the other hand, $Y_\pm(E)$ are nonanalytic functions and jumps steeply between $-\infty$ and $+\infty$ due to the divergence of the intergral in Eq. \eqref{evgpl}. Thus, both the ``$\pm$'' branches of Eq. \eqref{evgpl} have infinite numbers of roots in the regime $E>\omega_{11}$, which forms a continuous energy band. 

\begin{figure}[tbp]
\includegraphics[angle=-90,width=.55\columnwidth]{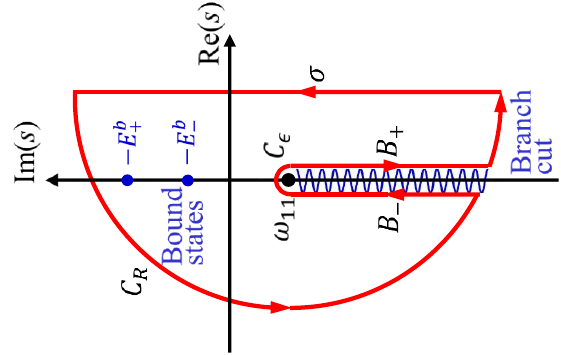}
\caption{Path of the contour integration in the complex plane of $s$ for calculating the inverse Laplace transform of $\tilde{c}_j(s)$. Two isolated poles at most are present in the imaginary axis of $s$.}\label{ctfg}
\end{figure}
Figure \ref{ctfg} shows the path of the contour integration to evaluate Eq. \eqref{invlps}. The poles from the continuous energy band $E$ form a branch cut. Two bound states at most formed in the imaginary axis give two isolated poles. According to the residue theorem, we have
\begin{eqnarray}
&&\lim_{\substack{\epsilon\rightarrow 0\\ R\rightarrow\infty}}\left[\int_{\sigma-iR}^{\sigma+iR}+\int_{C_R}+\int_{B_-}+\int_{C_\epsilon}+\int_{B_+}\right]\tilde{c}_j(s)e^{st}ds\nonumber\\
&&=2\pi i\sum_{p=\pm}\text{Res}(-iE^\text{b}_p),
\end{eqnarray}
where $\text{Res}(-iE^\text{b}_p)$ denotes the residue contributed by the $p$th bound state. From Jordan's lemma, the integrals along the arc paths $C_R$ and $C_\epsilon$ are negligible. Thus, the paths which have contribution to $c_j(t)$ are those around the branch cut and the two residues. We then have
\begin{eqnarray}
c_j(t)&=&\sum_{p=\pm}\text{Res}(-iE^\text{b}_p)\nonumber\\&&-{1\over2\pi i}\lim_{\epsilon\rightarrow 0}\left[\int_{B_-}+\int_{B_+}\right]\tilde{c}_j(s)e^{st}ds.\label{smsclt}
\end{eqnarray}
The integral along the imaginary axis of $s=iE$ of the branch cut in Eq. \eqref{smsclt} tends to zero in the long-time limit due to the out-of-phase interference of the components oscillating in continuously different frequencies $E$ in Eq. \eqref{smsclt}. Therefore, we are not bothered to calculate the integral in Eq. \eqref{smsclt} if only the steady-state form of $c_j(t)$ is concerned. Therefore, only the residues contributed by the isolated poles of the bound states are preserved in the long-time limit. The $p$th residue of $c_j(t)$ can be evaluated by the L'Hospital's rule. We have the following cases.
\begin{enumerate}
\item No bound state is present when $Y_{\pm}(\omega_{11})>\omega_{11}$. We have
\begin{equation}
c_j(\infty)=0,
\end{equation}which shows no qualitative difference from the Markovian result \eqref{mkvr}.
\item One bound state with eigenenergy $E_+^\text{b}$ is present when $Y_+(\omega_{11})<\omega_{11}$ and $Y_-(\omega_{11})>\omega_{11}$. Here, we have used the fact $Y_+(\omega_{11})<Y_-(\omega_{11})$. Then, $c_j(t)$ asymptotically tends to the residue of $-iE_+^\text{b}$. The residue is evaluated by the L'Hospital's rule as
\begin{eqnarray}
    ~~~~c_j(\infty)&=&\text{Res}(-iE_+^\text{b})\nonumber\\
    &=&\lim_{s\rightarrow -iE_+^\text{b}}{(s+iE_+^\text{b})e^{st}}\tilde{c}_j(s)\nonumber\\
    &=&Z_+e^{-iE_+^\text{b}t},\label{smobd}
\end{eqnarray} 
where $Z_\pm={1\over 2}[1+\int_0^\infty{J_0(\omega)\pm J_1(\omega)\over (\omega-E_\pm^\text{b})^2}d\omega]^{-1}$.
Equation \eqref{smobd} indicates that the exicted-state populations of the charger and the QB tends to a same constant in the long-time limit, which shows a qualitative difference from the Markovian result \eqref{mkvr}.  
\item Two bound states are present when $Y_{\pm}(\omega_{11})<\omega_{11}$. Then, $c_j(t)$ asymptotically tends to the summation of the two residues of $-iE_\pm^\text{b}$. The residues are evaluated as 
\begin{eqnarray}
    ~~~~c_j(\infty)&=&\sum_{p=\pm}\text{Res}(-iE_p^\text{b})\nonumber\\
    &=&\sum_{p=\pm}\lim_{s\rightarrow -iE_p^\text{b}}{(s+iE_p^\text{b})e^{st}}\tilde{c}_j(s)\nonumber\\
    &=&Z_+e^{-iE_+^\text{b}t}-(-1)^jZ_-e^{-iE_-^\text{b}t}.\label{tbsc}
\end{eqnarray}
Equation \eqref{tbsc} shows that, in dramatically different from the Markovian result \eqref{mkvr} and the single-bound-state result \eqref{smobd}, both the excited-state populations of the charger and the QB show a coherent interference between the two bound states in the long-time condition. 
\end{enumerate}
The above results demonstrate the dominate role of the feature of the energy spectrum of the total system in the non-Markovian dynamics of the open system. It supplies us an insightful guideline to suppress the destructive effect of the environment on the open system via engineering the energy-spectrum feature. 

\section{Full cycle of the scheme}
We see a widespread interest in designing different schemes of QB, including the ones without charger \cite{PhysRevLett.124.130601,PhysRevLett.125.040601} and the ones with charger \cite{PhysRevLett.120.117702,PhysRevLett.127.100601,PhysRevLett.129.130602} in recent years. An advantage of the QB schemes with a charger over the ones without a charger is that the charger is helpful to establish a QB-charger entanglement, which can act as a resource for generating nonzero extractable work \cite{PhysRevLett.129.130602}, and to induce a long-range interaction among different constitute subsystems of the QB, which can increase the charging
power \cite{PhysRevLett.120.117702}. Following this mainstream, we further demonstrate that the charger is helpful to overcome the aging of the QB and to form a non-interacting remote charging to the QB. 

\begin{figure}
    \includegraphics[width=\columnwidth]{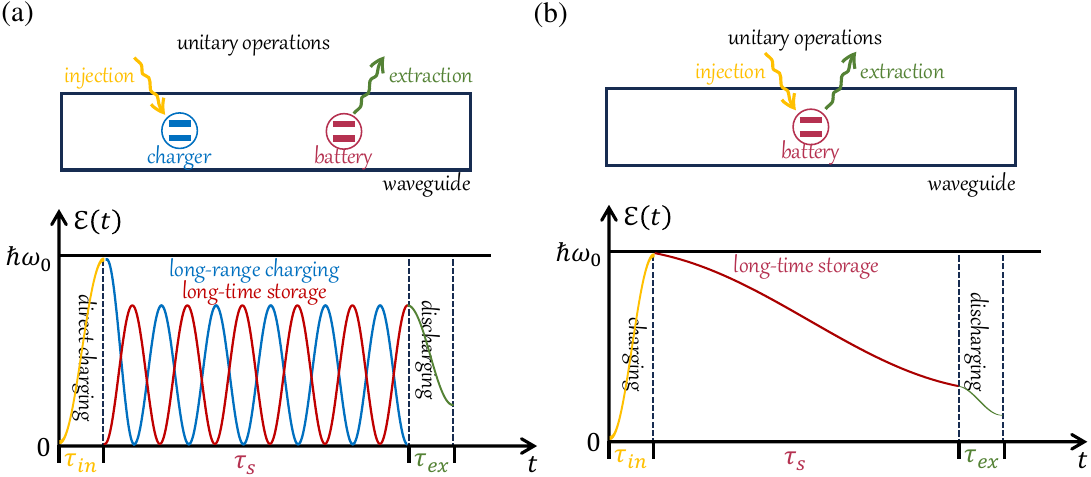}
    \caption{Upper panel of (a): QB protocol with a charger. The EMF in the waveguide as a common environment triggers a cooperative interplay between the QB and the charger manifesting the formation of two bound states such that the aging of the QB within the long-time storage process is overcome. Lower panel of (a): Full cycle of the scheme. The initialization process is within $t\in [0,\tau_{in})$. The combined charging and the storage processes are within $t\in[\tau_{in}, \tau_{in}+\tau_s)$, where the charging occurs in a reversible way and the energy loss of the QB is overcome governed by the two bound states. The energy extraction to the QB occurs within $t\in[\tau_{in}+\tau_s,\tau_{in}+\tau_s+\tau_{ex})$. Upper panel of (b): QB protocol without a charger. Lower panel of (b): Full cycle of the scheme. The QB is charged within $t\in[0,\tau_{in})$. The storage process is within $t\in[\tau_{in},\tau_{in}+\tau_{s})$. Without the protection of the waveguide induced cooperative charger-QB interplay, the QB energy is destroyed by the decoherence such that little energy can be extracted during the discharging process within $t\in[\tau_{in}+\tau_{s},\tau_{in}+\tau_{s}+\tau_{ex})$. }
    \label{fig:cycle}
\end{figure}
We design a QB scheme where the charger formed by a two-level system and a waveguide together act as a power station, see Fig. \ref{fig:cycle}(a). Our scheme uses the mediation role of the waveguide to induce an indirect coupling between the charger and the QB. Choosing a two-level system as the charger is because it is the minimal quantum system having the quantized interaction with the EMF in the waveguide. If a charging to the QB is needed, then the QB is taken to a distance not necessarily too near the charger and the charging via nearly coherent energy transfer can be automatically finished by the waveguide. In our scheme, the decoherence induced by the common electromagnetic environment in the waveguide triggers a cooperative interplay between the QB and the charger manifesting the formation of two bound states, which effectively counteracts the destructive effect of the decoherence on the QB. In contrast, if the charger is absent, see Fig. \ref{fig:cycle}(b), even we can charge the QB in some extra manner, the decoherence induced by the electromagnetic environment in the waveguide to the single QB would exhaust its energy, which is just the aging of QB. Then, neither anti-aging nor remote charging could be realized. 

The full cycle of our scheme is summarized as follows.
\begin{enumerate}
    \item Initialization. A $\pi$ laser pulse is applied on the charger to excite it from the ground state $|g\rangle$ to the excited state $|e\rangle$.
    \item Charging. If a charging to the QB is needed, then the QB is taken to the power station formed by the charger and waveguide. A remote charging to the QB is automatically done in a Rabi-like oscillation way, which is as perfect as the ideal case of charging via direct charger-QB interaction.  
    \item Storage. The energy is stored in the QB for an arbitrarily long time without energy loss because the charger-QB energy exchange is in a reversible manner. Thus, the energy-loss problem in the storage process generally encountered by the traditional QB schemes is efficiently avoided. 
    \item Discharging. If an energy supply from the QB is needed, the QB is taken out of the power station and an interaction between the QB and the target system is switched on. It causes the discharging of the QB. Although the QB in this process cannot be effectively protected by the waveguide from losing energy, its energy loss is small because this process occurs in a very short time scale.   
\end{enumerate}

The aging of QB generally occurs in the storage process because the time of the storage process is much longer than the one of the discharging process. In our scheme, the QB is kept in the waveguide to coherently exchange its energy with the charger as long as the energy extraction of the QB is not required. It means that our charging process and storage process are combined together for the purpose to effectively protect its energy. Because the energy exchange is in a reversible manner governed by the two bound states, the aging caused by the energy loss in the storage process is efficiently overcome. Our scheme supplies us with a feasible mechanism towards the direction of solving the long-standing aging problem of QB. 

\bibliography{QB}